\pdfoutput=1
\RequirePackage{ifpdf}
\ifpdf 
\documentclass[pdftex]{sigma}
\else
\documentclass{sigma}
\fi

\begin{document}

\newtheorem{thm}{Theorem}[section]
\newtheorem{prop}[thm]{Proposition}
\newtheorem{cor}[thm]{Corollary}
\newtheorem{cjt}[thm]{Conjecture}
\newtheorem{lem}[thm]{Lemma}

{\theoremstyle{definition}\newtheorem{de}[thm]{Definition}
\newtheorem{rem}[thm]{Remark}
\newtheorem{ex}[thm]{Example}}

\newcommand{\p}{\partial}
\newcommand{\ep}{\epsilon}
\def \la {\langle}
\def \ra{\rangle}
\def \dsum{\displaystyle\sum}

\allowdisplaybreaks

\renewcommand{\PaperNumber}{044}

\FirstPageHeading

\ShortArticleName{Commuting Dif\/ferential Operators of Rank 3 Associated to a Curve of Genus 2}

\ArticleName{Commuting Dif\/ferential Operators \\ of Rank 3 Associated to a Curve of Genus 2}

\Author{Dafeng ZUO~$^{\dag\ddag}$}

\AuthorNameForHeading{D.~Zuo}

\Address{$^\dag$~School of Mathematical Science,
University of Science and Technology of China,\\
\hphantom{$^\dag$}~Hefei 230026, P.R.~China}
\EmailD{\href{dfzuo@ustc.edu.cn}{dfzuo@ustc.edu.cn}}

\Address{$^\ddag$~Wu Wen-Tsun Key Laboratory of Mathematics, USTC, Chinese Academy of Sciences,\\
\hphantom{$^\ddag$}~P.R.~China}

\ArticleDates{Received March 12, 2012, in f\/inal form July 12, 2012; Published online July 15, 2012}

\Abstract{In this paper, we construct some examples of commuting dif\/ferential
opera\-tors~$L_1$ and~$L_2$ with rational coef\/f\/icients of rank 3 corresponding to a
curve of genus 2.}


\Classification{13N10; 14H45; 34L99; 37K20}

\section{Introduction}

The study of the commutation equation
\[
[L_1,L_2]=0
\]
of two scalar dif\/ferential operators
\begin{gather*}
 L_1=\frac{d^n}{dx^n}+\dsum_{i=0}^{n-1} f_i(x) \frac{d^i}{dx^i}\qquad
\mbox{and}\qquad L_2=\frac{d^{m}}{dx^{m}}+\dsum_{j=0}^{m-1} g_j(x) \frac{d^j}{dx^j},\qquad n<m,
\end{gather*}
is one  of the classical problems of the theory of ordinary
dif\/ferential equations.

Burchnall and Chaundy in \cite{BC1923,BC1928, BC1931}
have shown that {\it ``each pair of commuting operators $L_1$ and $L_2$ is connected by a
nontrivial polynomial algebraic relation $Q(L_1,L_2)=0$''.} The equation $Q(z,w)=0$
determines a smooth compact algebraic curve $\Xi$ of
f\/inite genus~$g$. For a generic point $P\in \Xi$, there exist common
eigenfunctions $\psi(x,P)$ on $\Xi$  such that
$L_1\psi=\lambda\psi$ and $L_2\psi=\mu\psi$. The dimension $l$ of the space
of these functions corresponding to $P\in \Xi$ is called the {\it rank} of
the commuting pair $(L_1,L_2)$.  For simplicity, in this paper we denote
``the commuting dif\/ferential operators of  rank $l$ corresponding to a curve
of genus $g$'' by {\it ``${(l,g)}$-operators''}.

Burchnall and Chaundy also made signif\/icant progress in solving the commutation
equation  for relatively prime orders $m$ and $n$. In this case, the
rank~$l$ equals to~$1$.  The study of this case was completed by Krichever
\cite{Kr1977, Kr1977-1}, who also obtained explicit formulas of
the function $\psi$ and the coef\/f\/icients of $L_1$ and $L_2$ in terms of the
Riemann $\Theta$-function. Let us remark that there are several papers related to
this case, for instance~\cite{Du, DMN,N1,P, Sh, W}.

But for high rank case i.e. $l>1$, it is  much more complicated. In \cite{Kr1978},
the problem of classifying $(l,g)$-operators was solved by reducing the
computation of the coef\/f\/icients to a Riemann problem.
In~\cite{KN1979, KN1978} I.M.~Krichever and S.P.~Novikov developed a method of defor\-ming
the Tyurin parameters on the moduli space of framed holomorphic bundles over algebraic curves.
By using this method, in certain cases the Riemann problem can be avoided and
they found all $(2,1)$-operators.
Let us remark that J.~Dixmier in \cite{D1968} also discovered an example of
$(2,1)$-operators with polynomial coef\/f\/icients. Furthermore, P.G.~Grinevich found
the condition of  $(2,1)$-operators with rational coef\/f\/icients \cite{G1982}.
S.P.~Novikov and P.G.~Grinevich \cite{NG1982} clarif\/ied the spectral data related to
formally self-adjoint $(2,1)$-operators.
In \cite{Mo1989} O.I.~Mokhov obtained all $(3,1)$-operators.
A.E.~Mironov in \cite{Mir2004,Mir2009} introduced a $\sigma$-invariance to simplify the
Krichever--Novikov system~\cite{KN1978} and  constructed some examples of
$(2,2)$-operators, $(2,4)$-operators with polynomial coef\/f\/icients and also in~\cite{Mir2005,Mir2011}
formally self-adjoint $(2,g)$-operators and $(3,g)$-operators. Recently, an interesting paper is
due to O.I.~Mokhov in~\cite{M2012} who constructed examples of $(2k,g)$-operators and $(3k,g)$-operators with polynomial
coef\/f\/icients for arbitrary genus $g$. For more related results, please see~\cite{Alex2,Alex1, KN1979,LP1995, LP1994,P,PM1992, PM1989}  and references therein.

The aim of this paper is to construct examples of commuting dif\/ferential operators~$L_1$ and~$L_2$ with
rational coef\/f\/icients of rank $3$ corresponding to a curve of genus~$2$, which is dif\/ferent from those
in~\cite{M2012}.

\section{The commuting operators of rank 3 and genus 2}

In this section we want to construct (3,2)-operators.  The f\/irst step is to use a $\sigma$-invariance, due to A.E.~Mironov \cite{Mir2004},
to simplify the Krichever--Novikov system \eqref{eq2.6}. The second step is to solve the simplif\/ied
system by making a crucial hypothesis
\begin{gather*}
 \gamma_1=\gamma,\qquad \gamma_2= a \gamma, \qquad \gamma_3=\bar{a}\gamma,\qquad a=\frac{-1+\sqrt{3}
{\bf i}}{2}.
\end{gather*}
 The last step is to construct the commuting dif\/ferential operators $L_1$ and $L_2$.

\subsection{The general principle}

Let $\Gamma$ be a curve of genus $2$ def\/ined in $\mathbb{C}^2$ by the equation
\[
 w^2=z^6+c_5z^5+c_4z^4+c_3z^3+c_2z^2+c_1z+c_0.
 \]
On the curve $\Gamma$, there is  a holomorphic involution
\[
 \sigma: \ \Gamma\rightarrow\Gamma\qquad \hbox{by}\quad \sigma(z,w)=(z,-w),
 \]
which has six f\/ixed ramif\/ication points. It induces an action on the space of function
by  $(\sigma f)(x,P)=f(x,\sigma(P))$.
Let us take $q=(0,\sqrt{c_0})\in \Gamma$.  For a generic point $P\in \Gamma$
there exist common eigenfunctions $\psi_j(x,P)$, $j=0,1,2$ with an essential singularity at $q$,
 of the opera\-tors~$L_1$ and~$L_2$. Without loss of generality, we assume that
 $\psi_j(x,P)$ are normalized by
\begin{gather*}
 \frac{d^i}{dx^i}\psi_j(x_0,P)=\delta_{ij},
 \end{gather*}
 where $x_0$ is a f\/ixed point.  Notice that on $\Gamma-\{q\}$, $\psi_j(x,P)$ are meromorphic and
have six simple poles at $P_1,\dots,P_6$ independent of~$x$. Let us consider the Wronskian matrix
\[
\vec{\Psi}(x,P;x_0)= \begin{pmatrix}
\psi_0& \psi_1&\psi_2\\
\psi_0'& \psi_1'&\psi_2'\\
\psi_0''& \psi_1''&\psi_2''
\end{pmatrix}
\]
of the vector-valued function $\vec{\Psi}(x,P;x_0)$, and
\begin{gather}
\vec{\Psi}_x\vec{\Psi}^{-1}= \begin{pmatrix}
0& 1&0\\
0& 0&1\\
\chi_0& \chi_1&\chi_2
\end{pmatrix},\label{eq2.2}
\end{gather}
where $\chi_j=\chi_j(x,P)$ are independent of $x_0$ and meromorphic functions on $\Gamma$ with six poles
at $P_1(x),\dots,P_6(x)$ coinciding with the poles of $\psi_j(x,P)$ at $x=x_0$. In a
neighborhood of $q$, the functions $\chi_j(x,P)$ have the form
\begin{gather}
   \chi_0(x,P)=k+w_0(x)+O\big(k^{-1}\big),\qquad
  \chi_1(x,P)=w_1(x)+O\big(k^{-1}\big), \nonumber\\
  \chi_2(x,P)=O\big(k^{-1}\big), \label{eq2.6}
\end{gather}
where $k^{-1}$ is a local parameter near $q$.
The expansion of $\chi_j$ in a neighborhood of the pole $P_i(x)$ has the form
\begin{gather}
 \chi_j(x,P)=-\frac{\gamma_i'(x)\alpha_{ij}(x)}{k-\gamma_i(x)}+d_{ij}(x)+O(k-\gamma_i(x)),\qquad
\alpha_{i2}=1,\label{eq2.7}
\end{gather}
where $k-\gamma_i(x)$ is a local parameter near $P_i(x)$ for $1\leq i \leq 6$ and $ 0\leq j\leq 2$.

\begin{lem}[\cite{Kr1977}] The parameters $\gamma_i(x)$, $\alpha_{ij}(x)$ and $d_{ij}(x)$, $
1\leq i \leq 6$, $0\leq j\leq 2$ satisfy
the system
\begin{gather}
{\rm Eq}[i,0]:=\alpha_{i0}(x)\alpha_{i1}(x)+\alpha_{i0}(x)d_{i2}(x)-\alpha'_{i0}(x)-d_{i0}(x)=0,\nonumber\\
{\rm Eq}[i,1]:= \alpha_{i1}(x)^2-\alpha_{i0}(x)+\alpha_{i1}(x)d_{i2}(x)-\alpha'_{i1}(x)-d_{i1}(x)=0.
 \label{eq2.8}
\end{gather}
\end{lem}

\subsection[Explicit forms of $\chi_j(x,P)$]{Explicit forms of $\boldsymbol{\chi_j(x,P)}$}

In this subsection,  we discuss explicit forms of $\chi_j(x,P)$
corresponding to the curve $\Gamma$ def\/ined by  $w^2=1+c_3z^3+c_4z^4+z^6$.
In order to do this, we assume that
\begin{gather}
 \sigma \chi_2(x,P)=\chi_2(x,P),\qquad \sigma P_s(x)=P_{s+3}(x),\qquad s=1,2,3,\label{eq2.10}
\end{gather}
and
\begin{gather}
\gamma_1=\gamma,\qquad \gamma_2= a \gamma, \qquad \gamma_3=\bar{a}\gamma,\qquad
a=\frac{-1+\sqrt{3}\,{\bf i}}{2}.\label{eq2.11}
\end{gather}

\begin{thm}\label{thm2.2}
Let $\gamma$ be a solution of
\begin{gather} 1+c_3\gamma^3+\gamma^6-6 (-3)^{\frac{1}{4}}c_4^{\frac{1}{4}}\gamma'^{\frac{3}{2}}=0,
\label{eq2.9}
\end{gather}
then functions $\chi_0$, $\chi_1$, $\chi_2$ are given by the formulas
\begin{gather}
 \chi_2(x,P)=-\sum_{s=1}^3\frac{\gamma_s'}{z-\gamma_s}-\sum_{s=1}^3\frac{\gamma_s'}{\gamma_s}
=\frac{3z^3\gamma'}{\gamma^4-z^3\gamma},\nonumber\\
 \chi_1(x,P)=\tau_1-\sum_{s=1}^3\frac{G_s\gamma_s'}{z-\gamma_s}+
\frac{w(z)h_1}{2(z-\gamma_1)(z-\gamma_2)(z-\gamma_3)},\label{eqz2.11}\\
 \chi_0(x,P)=\frac{\tau_0}{2}+\frac{1}{2z}-\sum_{s=1}^3\frac{H_s\gamma_s'}{z-\gamma_s}-
\frac{w(z)(\gamma_1\gamma_2\gamma_3+z h_0)}{2z(z-\gamma_1)(z-\gamma_2)(z-\gamma_3)}
,\nonumber
\end{gather}
with $G_s$, $H_s$, $\tau_0$,  $\tau_1$ defined in \eqref{eq2.14}--\eqref{eq2.21}.
\end{thm}

\begin{proof} By using the $\sigma$-invariance of $\chi_2(x,P)$, we know
\[
\gamma_s(x)=\gamma_{s+3}(x), \qquad d_{s2}(x)=d_{s+3,2}(x), \qquad s=1,2,3.
\]
According to the properties of $\chi_j(x,P)$ in~\eqref{eq2.6},~\eqref{eq2.7} and~\eqref{eq2.10},
we could assume that the functions $\chi_j(x,P)$ are of the form in~\eqref{eqz2.11} with
 unknown functions  $G_s=G_s(x)$, $H_s=H_s(x)$, $\tau_r=\tau_r(x)$ and
$h_r=h_r(x)$ for $s=1,2,3$ and $r=0,1$.

Substituting \eqref{eq2.11} into \eqref{eqz2.11}, we have
\[
\chi_2(x,P)=\frac{3z^3\gamma'}{\gamma^4-z^3\gamma},
\]
 which yields that
\[
 d_{i2}=-\frac{2\gamma'}{\gamma},\qquad i=1,\dots,6.
 \]
For simplicity we use the following notations
\begin{gather*}
a_1=1,\qquad a_2=a, \qquad a_3=\bar{a}, \\ a_{s+3}=a_s, \qquad G_{s+3}=G_s, \qquad H_{s+3}=H_s, \qquad s=1,2,3.
\end{gather*}
It follows from \eqref{eq2.7} that
\begin{gather*}
  \alpha_{s0}=H_s+\frac{w(a_s\gamma)h_0}{6\gamma^2\gamma'}+\frac{a_s^2w(a_s\gamma)}{6\gamma'},\qquad
\alpha_{s1}=G_s-\frac{w(a_s\gamma)h_1}{6\gamma^2\gamma'}, \\
  d_{s0}=\frac{\tau_0}{2}+\frac{a_s^2}{2\gamma}+\frac{\gamma'\dsum_{m=1}^2(1-a_s^2a_{s+m})H_{s+m}}{3\gamma}\\
\hphantom{d_{s0}=}{} +\frac{(h_0+2(a_s\gamma)^2)w(a_s\gamma)-(h_0+(a_s\gamma)^2)a_s\gamma w'(a_s\gamma)}{6\gamma^3}, \\
 d_{s1}=\tau_1-\frac{G_{s+1}\gamma'}{(a_s-a_{s+1})\gamma}-\frac{G_{s+2}\gamma'}{(a_s-a_{s+2})\gamma}
+\frac{(a_s\gamma w'(a_s\gamma)-w(a_s\gamma))h_1}{6\gamma^3}, \\
  \alpha_{s+3,r}= \sigma \alpha_{sr},\qquad  d_{s+3,1}=\sigma d_{s1},\qquad r=0,1,\qquad
s=1,2,3.
\end{gather*}
By substituting $\alpha_{ij}$ and $d_{ij}$ into \eqref{eq2.8}, we get twelve equations
\[
{\rm Eq}[i,0]=0,\qquad {\rm Eq}[i,1]=0, \qquad i=1,\dots,6.
\]
We now try to solve these equations. Firstly, it follows from
\[
{\rm Eq}[s+3,1]-{\rm Eq}[s,1]=0,\qquad s=1,2,3
\] that
\begin{gather}
 G_s=\frac{h_1'-h_0-(a_s\gamma)^2}{2h_1}+\frac{\gamma'}{2\gamma}-\frac{\gamma''}{2\gamma'},
\qquad s=1,2,3.\label{eq2.14}
\end{gather}
By using \eqref{eq2.14} and  ${\rm Eq}[s+3,0]-{\rm Eq}[s,0]=0$, we get
\begin{gather}
 H_s=\frac{(h_0+(a_s\gamma)^2)h_1'-2h_0'-7a_s^2\gamma\gamma'}{2h_1}
-\frac{(h_0+(a_s\gamma)^2)^2}{2h_1^2}  \\
\hphantom{H_s=}{}  -\frac{h_0\gamma'}{2h_1\gamma}+\frac{(h_0+(a_s\gamma)^2)\gamma''}{2h_1\gamma'},
\qquad s=1,2,3.  \label{AD1}
\end{gather}
Furthermore, by solving
\[
{\rm Eq}[s+3,1]+{\rm Eq}[s,1]=0,\qquad {\rm Eq}[s+3,0]+{\rm Eq}[s,0]=0,
\]
we have
\begin{gather*}
{\rm Neq} [s,1]:=-\tau_1+\frac{h_0^2+6h_0(a_s\gamma)^2+6a_s\gamma^4-6h_0h_1'
-6(a_s\gamma)^2h_1'+3h_1'^2}{4h_1^2} \\
\hphantom{{\rm Neq} [s,1]:=}{}
+ \frac{3h_0'-h_1''+9a_s^2\gamma\gamma'}{2h_1}+\frac{3h_0\gamma'-2h_1'\gamma'}{2h_1\gamma}
+\frac{\gamma''}{2\gamma}
+\frac{\gamma'''}{2\gamma'} \\
\hphantom{{\rm Neq} [s,1]:=}{}
 -\frac{3\gamma'^2}{4\gamma^2}-\frac{\gamma''^2}{4\gamma'^2}-\frac{h_1\gamma''}{2h_1\gamma'}
+\frac{h_1^2w(a_s\gamma)}{36\gamma^4\gamma'^2}=0,\qquad s=1,2,3,
\end{gather*}
and
\begin{gather*}
{\rm Neq}[s,0]:=-\tau_0-\frac{a_s^2}{\gamma}+\frac{4h_0''+16a_s^2\gamma\gamma'+21(a_s\gamma)^2}{2h_1} \\
\hphantom{{\rm Neq}[s,0]:=}{}
  +\frac{(3h_0'+9a_s\gamma\gamma'-h_1'')(h_0+(a_s\gamma)^2)-4h_0'h_1'-13a_s^2\gamma\gamma'h_1'}{h_1^2} \\
\hphantom{{\rm Neq}[s,0]:=}{}
+ \frac{(h_0+(a_s\gamma)^2)^3-6h_1'(h_0+(a_s\gamma)^2)^2+5h_1'^2(h_0+(a_s\gamma)^2)}{2h_1^3}
 \\
\hphantom{{\rm Neq}[s,0]:=}{}
+ \frac{6h_0'\gamma'-h_0\gamma''}{h_1\gamma}
+\frac{(3h_0^2-4h_0h_1')\gamma'}{h_1^2\gamma}-\frac{(h_0+(a_s\gamma)^2)\gamma'''}{h_1\gamma'}
 \\
\hphantom{{\rm Neq}[s,0]:=}{}
+\frac{(h_0+(a_s\gamma)^2)h_1'\gamma''}{h_1^2\gamma'}
+\frac{(h_0+(a_s\gamma)^2)\gamma''^2}{2h_1\gamma'^2}+\frac{3h_0\gamma'^2}{2h_1\gamma^2} \\
\hphantom{{\rm Neq}[s,0]:=}{}
 -\frac{(h_0+(a_s\gamma)^2)h_1w(a_s\gamma)}{18\gamma^4\gamma'^2},\qquad s=1,2,3.
\end{gather*}
Let us remark that we have reduced twelve equations to six equations
\[
{\rm Neq}[s,0]=0,\qquad {\rm Neq}[s,1]=0,\qquad s=1,2,3,
\]
with four unknown functions $\tau_1$, $\tau_0$, $h_1$ and $h_0$.

Let us take
\begin{gather}
 h_1=i(-3)^\frac{3}{4}c_4^{-\frac{1}{4}}\gamma\sqrt{\gamma'},\qquad
h_0=\frac{i(-3)^\frac{3}{4}(\gamma\gamma''-4\gamma'^2)}{2c_4^{\frac{1}{4}}\sqrt{\gamma'}}. \label{AD2}
\end{gather}
From ${\rm Neq}[1,1]=0$, we get
\begin{gather}
 \tau_1=\frac{4\gamma'^2-9\gamma\gamma''}{2\gamma^2}+\frac{4\gamma'\gamma'''-3\gamma''^2}
{4\gamma'^2}+\frac{i(\gamma^3-1)^2}{4\sqrt{3c_4}\gamma^2\gamma'}.  \label{eq2.19}
\end{gather}
By using \eqref{eq2.19}, we conclude that  ${\rm Neq}[2,1]=0$ and ${\rm Neq}[3,1]=0$ always hold true.

From the equation ${\rm Neq}[1,0]=0$, we obtain
\begin{gather}
\tau_0=\frac{i(\gamma^3-1)^2}{\sqrt{3c_4}\gamma^3}-\frac{1}{\gamma}-
\frac{i(\gamma^3-1)^2\gamma''}{4\sqrt{3c_4}\gamma^2\gamma'^2}-
\frac{2 i(-3)^{\frac{3}{4}}c_4^{\frac{3}{4}}\gamma^3}{27\gamma'^{\frac{3}{2}}}-
\frac{i (-3)^{\frac{3}{4}}(\gamma^3-1)^2}{18c_4^{\frac{1}{4}}\gamma\gamma'^{\frac{3}{2}}}\nonumber\\
\phantom{\tau_0=}{}
-\frac{3\gamma'''}{\gamma}+\frac{10\gamma'\gamma''}{\gamma^2}-
\frac{4\gamma'^3}{\gamma^3}+\frac{\gamma^{(4)}}{\gamma'}-\frac{5\gamma''\gamma'''}{2\gamma'^2}
+\frac{3\gamma''^3}{2\gamma'^3}-\frac{3\gamma''^2}{\gamma\gamma'}. \label{eq2.21}
\end{gather}
By using \eqref{eq2.21},  both ${\rm Neq}[2,0]=0$ and ${\rm Neq}[3,0]=0$ reduce to the same equation
\[
 1+c_3\gamma^3+\gamma^6-6 (-3)^{\frac{1}{4}}c_4^{\frac{1}{4}}\gamma'^{\frac{3}{2}}=0,
\]
which is exactly the equation~\eqref{eq2.9}. Thus we complete the proof of the theorem.
\end{proof}

 Generally, solutions of \eqref{eq2.9} are
 not useful for us to construct $(3,2)$-operators with ``good'' coef\/f\/icients.
But when we choose $c_3=2$ or $-2$, there are rational solutions. In what follows let us suppose
\[
 c_3=-2,\qquad c_4=-\frac{\epsilon^4}{3888},\qquad \epsilon<0.
 \]
The equation \eqref{eq2.9} is rewritten as
\begin{gather}
 1-2 \gamma^3+\gamma^6+\epsilon\gamma'^{\frac{3}{2}}=0. \label{eq2.20}
\end{gather}
It is easy to check that when $(x+s_0)^3+\epsilon^2>0$,
\[
 \gamma=\frac{x+s_0}{((x+s_0)^3+\epsilon^2)^{\frac{1}{3}}}, \qquad  s_0\in \mathbb{C}
 \]
is a solution of \eqref{eq2.20}. Without loss of generality, we set $s_0=0$. In this case we
would like to write $\gamma=\gamma(x;\epsilon)$.
As a corollary of  Theorem~\ref{thm2.2}, we have

\begin{cor}
Let $\gamma(x;\epsilon)=\frac{x}{(x^3+\epsilon^2)^{\frac{1}{3}}}$ be a solution of \eqref{eq2.20}.
Then we have
\begin{gather}
 \chi_0(x,P)=\frac{1}{2z}-\frac{x^3(\ep^2+x^3)}{5832}+\frac{10(z^3-1)}{\kappa}+\frac{\ep^2 x^3z}{216\kappa}\nonumber\\
  \phantom{\chi_0(x,P)=}{} -\frac{108 w(z)+\ep^2z^2}{6\kappa}
-\frac{x^3w(z)}{2\kappa z}+\frac{16\ep^2z^3}{\kappa x^3},\nonumber\\
 \chi_1(x,P)=\frac{132\ep^2z^3-x^3[204-204z^3+108 w(z)+\ep^2z^2]}{12x^2\kappa},\qquad
  \chi_2(x,P)=-\frac{3\ep^2z^3}{x\kappa},\label{eq2.22}
\end{gather}
where $\kappa=(\ep^2+x^3)z^3-x^3$ and $w(z)=\sqrt{1-2z^3-\frac{\ep^2}{3888}z^4+z^6}$.
\end{cor}

 By using \eqref{eq2.22}, let us expand $\chi_j(x,P)$ in a neighborhood of $z=0$
\begin{gather*}
 \chi_0(x,P)=\frac{1}{z}+\zeta_1-\frac{\ep^2}{216}z+\frac{2\ep^2}{3x^2}z^2+O\big(z^3\big), \\
 \chi_1(x,P)=\zeta_2+\frac{\ep^2}{12x^2}z^2+O\big(z^3\big),\qquad
 \chi_2(x,P)=\frac{3\ep^2}{x^4}+O\big(z^4\big),
\end{gather*}
where
\begin{gather}
 \zeta_1=\dfrac{28}{x^2}-\dfrac{\ep^2x^3+x^6}{5832} \qquad \mbox{and}\qquad \zeta_2=\dfrac{26}{x^2}.
\label{NZ1}
\end{gather}

\subsection{Commuting dif\/ferential operators of rank 3}

Let $\Gamma$ be a smooth curve of genus $2$ def\/ined
by the equation
\begin{gather}
 w^2=1-2z^3-\frac{\ep^4}{3888}z^4+z^6\label{AD4}
\end{gather}
on the $(z,w)$-plane.

\begin{thm}\label{Main} The operator $L_1$   corresponding to the meromorphic function
\[
\lambda=\dfrac{1+w(z)}{2z^3}-\dfrac{1}{2}
\]  on $\Gamma$ with the
unique pole at $q=(0,1)$ and $L_1\psi=\lambda\psi$
has the form
\begin{gather}
 L_1=\frac{d^9}{dx^9}+\dsum_{n=0}^7 f_n \frac{d^n}{dx^n}, \label{zuo2.17}
 \end{gather}
where
\begin{gather}
 f_0=\frac{152}{243}-\frac{58240}{x^9}-\frac{55 \ep^2}{243 x^3}-\frac{37 \ep^4 x^3}{11337408}+
\frac{115 \ep^2 x^6}{11337408}+\frac{37 x^9}{1417176} \nonumber\\
\hphantom{f_0=}{}
+\frac{\ep^6 x^9}{198359290368}+\frac{\ep^4 x^{12}}{66119763456}+\frac{\ep^2 x^{15}}{66119763456}
+\frac{x^{18}}{198359290368},\nonumber\\
 f_1=\frac{58240}{x^8}+\frac{55\ep^2}{243 x^2}-
\frac{152 x}{243}+\frac{5 \ep^4 x^4}{5668704}+\frac{2 \ep^2 x^7}{177147}+\frac{17 x^{10}}{1417176},
\nonumber\\
f_2=-\frac{43200}{x^7}+\frac{26\ \ep^2}{243 x}-\frac{73 x^2}{243}+\frac{\ep^4 x^5}{1259712}
+\frac{\ep^2 x^8}{419904}+\frac{x^{11}}{629856},
\nonumber \\
 f_3=-\frac{143 \ep^2}{1944}+\frac{19120}{x^6}+\frac{79 x^3}{486}+
\frac{\ep^4 x^6}{11337408}+\frac{\ep^2 x^9}{5668704}+\frac{x^{12}}{11337408},
\nonumber \\
 f_4=-\frac{4800}{x^5}-\frac{2 \ep^2 x}{243}+\frac{16 x^4}{243},
  \qquad f_5=-\frac{24}{x^4}+\frac{\ep^2 x^2}{216}+\frac{x^5}{108},
\nonumber \\
 f_6=\frac{384}{x^3}+\frac{\ep^2 x^3}{1944}+\frac{x^6}{1944},\qquad
 f_7=-\frac{78}{x^2}.
\label{zuo2.18}
\end{gather}
\end{thm}

\begin{proof} By using \eqref{eq2.2}, we have
\begin{gather}
\psi_j'''(x,P)=\chi_2(x,P)\psi_j''(x,P)+\chi_1(x,P)\psi_j'(x,P)+\chi_0(x,P)\psi_j(x,P).\label{eq2.27}
\end{gather}
It follows from \eqref{eq2.27} that the equation $L_1\psi_j=\lambda(z)\psi_j$ can be rewritten as
\begin{gather}
 Q_0(x,z)\psi_j(x,P)+Q_1(x,z)\psi_j'(x,P)+Q_2(x,z)\psi_j''(x,P)=\lambda(z)\psi_j.\label{eq2.28}
\end{gather}
According to the independence of $\chi_0(x,P)$, $\chi_1(x,P)$ and $\chi_2(x,P)$ at $x=x_0$, we conclude that
the system \eqref{eq2.28} is equivalent to three equations
\[
Q_0(x,z)=\lambda(z),\qquad Q_1(x,z)=0,\qquad Q_2(x,z)=0.
\]
By expanding $Q_j(x,z)$ at $z=0$, we have
\[
 0=Q_j(x,z)-\delta_j^0\lambda(z)=Q_{j,-2}\frac{1}{z^{2}}+Q_{j,-1}\frac{1}{z}+Q_{j0}+O(z).
\]
Then by solving $Q_{j,-s}=0$ for $s,j=0,1,2$, we get the coef\/f\/icients of $L_1$ given by
\begin{gather*}
  f_0=-1-\zeta_1^3-\frac{4\ep^2+3}{2x^3}-\zeta_2 \zeta_1' \zeta_2'-
-\zeta_2^2 \zeta_1''+6 \zeta_1' \zeta_1''+3 \zeta_1'' \zeta_2''+
3 \zeta_2' \zeta_1''' \\
\hphantom{f_0=}{} +\zeta_1 \left(-\frac{\ep^2}{72}-3 \zeta_2 \zeta_1'+3 \zeta_1'''\right)+
\zeta_1' \zeta_2'''+2 \zeta_2 \zeta_1^{(4)}-\zeta_1^{(6)}, \\
  f_1=-\frac{1}{4x^2}+6 \zeta_1'^2+9 \zeta_1 \zeta_1''+
12 \zeta_2' \zeta_1''+9 \zeta_1' \zeta_2''+3 \zeta_2''^2-
\zeta_2^2 (3 \zeta_1'+\zeta_2'')+3 \zeta_1 \zeta_2''' \\
\hphantom{f_1=}{}
 +4 \zeta_2' \zeta_2'''+\zeta_2 \left(-\frac{\ep^2}{72}-3 \zeta_1^2-3 \zeta_1 \zeta_2'-
\zeta_2'^2+9 \zeta_1'''+2 \zeta_2^{(4)}\right)-6 \zeta_1^{(5)}-\zeta_2^{(6)},   \\
 f_2=3 \big[{-}\zeta_2^2 \zeta_2'+5 \zeta_1' \zeta_2'+
5 \zeta_2' \zeta_2''-\zeta_1\zeta_2^2+3\zeta_1 (\zeta_1'+\zeta_2'')+
\zeta_2 (5 \zeta_1''+3 \zeta_2''')-5 \zeta_1^{(4)}-2 \zeta_2^{(5)}\big], \\
  f_3=\frac{\ep^2}{72}+3 \zeta_1^2-\zeta_2^3+9 \zeta_1 \zeta_2'+9 \zeta_2'^2+
3 \zeta_2 (4 \zeta_1'+5 \zeta_2'')-21 \zeta_1'''-15 \zeta_2^{(4)},  \\
  f_4=15 \zeta_2 \zeta_2'-\zeta_2 (2 (-3 \zeta_1-9 \zeta_2')+21 \zeta_2')-
18 \zeta_1''-21 \zeta_2''',  \\
  f_5=3 \zeta_2^2-9 \zeta_1'-18 \zeta_2'',\qquad f_6=-3 \zeta_1-9 \zeta_2',\qquad f_7=-3 \zeta_2.
\end{gather*}
By substituting $\zeta_1$ and $\zeta_2$ in  \eqref{NZ1} into the above formula, we obtain
explicit expressions of $f_j$ in~\eqref{zuo2.18}. \end{proof}

Next we want to look for a $12^{\rm th}$-order dif\/ferential operator
\begin{gather}
 L_2=\dfrac{d^{12}}{dx^{12}}+\dsum_{m=0}^{10} g_m \dfrac{d^m}{dx^m},\label{AD10}
 \end{gather}
such that $[L_1,L_2]=0$. Let us sketch out our ideas and omit tedious computations.
 The commutation equation $[L_1,L_2]=0$ is written as
\begin{gather}
 0=\left[\frac{d^{9}}{dx^{9}}+\dsum_{n=0}^{7} f_n \frac{d^n}{dx^n}, \frac{d^{12}}{dx^{12}}
 +\dsum_{m=0}^{10} g_m \frac{d^m}{dx^m}\right]= \dsum_{k=0}^{18}W_k(f,g)\frac{d^k}{dx^k}, \label{AD5}
 \end{gather}
which yields that
\[ W_k(f,g)=0,\qquad k=0,\dots,18.
 \]
By using eleven equations $W_k(f,g)=0$, $k=8,\dots,18$, we could obtain explicit forms of
$g_m=h_m(x;\rho_0,\dots,\rho_{10-m})+\rho_{11-m}$ with integral constants $\rho_{11-m}$.
The last eight equations will determine some integral constants. For simplicity, we take
all arbitrary parameters to be zero, and then obtain all coef\/f\/icients $g_j$ as follows
\begin{gather}
 g_0=\frac{45660160}{x^{12}}-\frac{4928 \ep^2}{729 x^6}-\frac{20048}{729 x^3}-\frac{605 \ep^2 x^3}{708588}
 +\frac{4553 x^6}{708588}+\frac{79
\ep^6 x^6}{99179645184}\nonumber\\
\hphantom{g_0=}{} +\frac{269 \ep^4 x^9}{16529940864} +\frac{683 \ep^2 x^{12}}{16529940864}
+\frac{\ep^8 x^{12}}{1156831381426176}+\frac{661
x^{15}}{24794911296}\nonumber\\
\qquad+\frac{\ep^6 x^{15}}{289207845356544}+
\frac{\ep^4 x^{18}}{192805230237696}+\frac{\ep^2 x^{21}}{289207845356544}+\frac{x^{24}}{1156831381426176},
\nonumber\\
g_1=-\frac{45660160}{x^{11}}+\frac{4928 \ep^2}{729 x^5}+\frac{20048}{729 x^2}
-\frac{203 \ep^4 x}{2834352}+\frac{1691 \ep^2
x^4}{2834352}+\frac{7111 x^7}{708588} \nonumber\\
\hphantom{g_1=}{}  +\frac{55 \ep^6 x^7}{49589822592}+\frac{127 \ep^4 x^{10}}{16529940864}
+\frac{217 \ep^2 x^{13}}{16529940864}+\frac{325
x^{16}}{49589822592}, \nonumber\\
 g_2=\frac{27758080}{x^{10}}-\frac{182 \ep^2}{27 x^4}+\frac{296}{9 x}
 -\frac{413 \ep^4 x^2}{5668704}+\frac{4339 \ep^2
x^5}{2834352}+\frac{6595 x^8}{1417176} \nonumber\\
\hphantom{g_2=}{}
+\frac{\ep^6 x^8}{3673320192}+\frac{\ep^4 x^{11}}{918330048}+\frac{5 \ep^2 x^{14}}{3673320192}
+\frac{x^{17}}{1836660096},\nonumber\\
g_3=-\frac{5992}{729}-\frac{11567360}{x^9}+\frac{1028 \ep^2}{729 x^3}+\frac{25 \ep^4 x^3}{1417176}
+\frac{457 \ep^2 x^6}{708588}+\frac{1393
x^9}{1417176} \nonumber\\
\hphantom{g_3=}{}
+\frac{\ep^6 x^9}{49589822592}+\frac{\ep^4 x^{12}}{16529940864}+\frac{\ep^2 x^{15}}{16529940864}
+\frac{x^{18}}{49589822592} ,\nonumber\\
 g_4=\frac{3395840}{x^8}+\frac{271 \ep^2}{243 x^2}-\frac{2834 x}{243}+\frac{193 \ep^4 x^4}{11337408}
 +\frac{317 \ep^2
x^7}{2834352}+\frac{307 x^{10}}{2834352}, \nonumber\\
 g_5=-\frac{693504}{x^7}-\frac{13 \ep^2}{243 x}+\frac{221 x^2}{243}+\frac{\ep^4 x^5}{314928}
 +\frac{\ep^2 x^8}{104976}+\frac{x^{11}}{157464},\nonumber\\
g_6=-\frac{167 \ep^2}{972}+\frac{86464}{x^6}+\frac{316 x^3}{243}+\frac{\ep^4 x^6}{5668704}
+\frac{\ep^2 x^9}{2834352}+\frac{x^{12}}{5668704},  \nonumber\\
 g_7=-\frac{672}{x^5}+\frac{\ep^2 x}{486}+\frac{109 x^4}{486},\qquad
 g_8=-\frac{2856}{x^4}+\frac{\ep^2 x^2}{108}+\frac{x^5}{54},  \nonumber\\
 g_9=\frac{824}{x^3}+\frac{\ep^2 x^3}{1458}+\frac{x^6}{1458},
 \qquad  g_{10}=-\frac{104}{x^2}.
 \label{zuo2.19}
 \end{gather}

\begin{rem}
By analogy with the process of getting $f_j$ in \eqref{zuo2.18},
we could obtain the above $g_j$ in~\eqref{zuo2.19} by choosing another meromorphic
function with a unique pole of order $4$ at $z=0$ on~$\Gamma$
\[
 \mu(z)=\dfrac{1+w(z)}{2z^4}-\frac{1}{2z}.
 \]
 \end{rem}

 \begin{rem} One could f\/ind another operator $L_3$ of order $15$ from $[L_1,L_3]=0$. Furthermore
 as in \cite{Mir2004}, the  commutative ring of dif\/ferential operators generated by $L_1$, $L_2$ and
 $L_3$ is isomorphic to the ring of meromorphic functions on $\Gamma$ with the pole at $q=(0,1)$.
 \end{rem}

 \subsection[The corresponding Burchnall-Chaundy curve]{The corresponding Burchnall--Chaundy curve}

 According to the Burchnall--Chaundy's correspondence in \cite{BC1923,BC1928, BC1931},
 for each pair of commuting operators $L_1$ and $L_2$
 there is a Burchnall--Chaundy curve def\/ined by a minimal nontrivial
 polynomial $Q(z,w)=0$ such that $Q(L_1,L_2)=0$ (or $Q(L_2,L_1)=0$). Obviously,
the above curve~$\Gamma$ def\/ined by \eqref{AD4} is not the Burchnall--Chaundy curve for $L_1$ and $L_2$ given in~\eqref{zuo2.17} and~\eqref{AD5}.
 Actually the corresponding Burchnall--Chaundy curve $\tilde{\Gamma}$
is given by
\[
 w^3-\frac{\ep^4}{15552}w^2=z^4+z^3,
 \]
that is to say,
\[
 L_2^3-\frac{\ep^4}{15552}L_2^2=L_1^4+L_1^3.
 \]
 The curve $\tilde{\Gamma}$  has a cuspidal singularity at $(0,0)$.  The operators $L_1$
 and $L_2$ correspond to those meromorphic functions on $\Gamma$
\[
  \lambda=\dfrac{1+w(z)}{2z^3}-\frac{1}{2},\qquad \mu=\dfrac{1+w(z)}{2z^4}-\frac{1}{2z}
  \]
def\/ining a birational equivalence
\[
 \pi: \ \Gamma \rightarrow  \tilde{\Gamma}, \qquad \pi(z,w)=(\lambda,\mu).
 \]
The inverse image of the cuspidal point is the point $\sigma(q)$, where $q=(0,1)\in \Gamma$.
In order to make~$\pi$ to be a morphism, we must complement $\tilde{\Gamma}$ at inf\/inity by a
cuspidal point of the type $(3,4)$, then its inverse image is the point~$q$.

\section{Concluding remarks}

In summary by using a $\sigma$-invariance to simplify the
Krichever--Novikov system, we have constructed a pair of commuting
dif\/ferential operators $L_1$ in \eqref{zuo2.17} and $L_2$ in \eqref{AD10} of rank $3$ with
rational coef\/f\/icients corresponding to the singular curve $\tilde{\Gamma}$,
which is birationally equivalent to the smooth curve $\Gamma$ of genus $2$.

Let us remark that all of coef\/f\/icients of $L_1$ and
$L_2$ are polynomials with respect to the
parameter $\ep$. So if we take
\[
 \mathcal{L}_1=\lim_{\ep \to 0}L_1,\qquad \mathcal{L}_2=\lim_{\ep \to 0}L_2,
 \]
then
\[
[\mathcal{L}_1,\mathcal{L}_2]=0,\qquad \mathcal{L}_2^3=\mathcal{L}_1^4+\mathcal{L}_1^3.
\]
More precisely,  we have
\[
\mathcal{L}_1=\mathcal{L}^3-1,\qquad \mathcal{L}_2=\mathcal{L}^4-\mathcal{L},
\]
where
\[
\mathcal{L}=\dfrac{d^3}{dx^3}-\dfrac{26}{x^2}\dfrac{d}{dx}-\dfrac{28}{x^3}+\dfrac{x^6}{5832}.
\]
So, when $\ep=0$ this is a trivial example.

How about the case $\ep\ne 0$?  Let us comment that in this case,
by a direct verif\/ication there is not such kind of~$\mathcal{L}$ of order~$3$ commuting
with~$L_1$ and~$L_2$.  Furthermore, according to the result in~\cite{W}, any rank one operator with
rational coef\/f\/icients whose second highest coef\/f\/icient is zero has the property that the limit as
$x$ goes to $\infty$ of the coef\/f\/icients is zero. So, for example, the absence of a
$\frac{d^{11}}{dx^{11}}$ term in~$L_2$ and the~$x^6$ in the coef\/f\/icient of its
$\frac{d^9}{dx^9}$ term which means that~$L_2$ is not a rank~$1$ operator.

\subsection*{Acknowledgments}

The author is grateful to Andrey E.~Mironov
for bringing 
the attention to 
this project and helpful discussions. The author
also thanks referees' suggestions and Alex Kasman for pointing some errors in the f\/irst
version of this paper, 
Qing Chen and Youjin Zhang for their constant supports.
This work is supported by ``PCSIRT" and the Fundamental Research
Funds for the Central Universities (WK0010000024) and NSFC (No.~10971209)
and SRF for ROCS, SEM.

\pdfbookmark[1]{References}{ref}
\LastPageEnding

\end{document}